\documentclass[11pt]{article}
\usepackage{natbib,cospar}

\usepackage{graphicx}

\newcommand{\CIV}{C~{\sc iv}}

\newcommand{\MgII}{Mg~{\sc ii}}

\newcommand{\kms}{\hbox{km~s$^{-1}$}}
\newcommand{\flux}{\hbox{erg~cm$^{-2}$~s$^{-1}$}}

\newcommand{\cmsq}{\hbox{cm$^{-2}$}}

\newcommand{\gtrsim}{\lower 2pt \hbox{$\, \buildrel {\scriptstyle >}\over {\scriptstyle\sim}\,$}}
\newcommand{\lesssim}{\lower 2pt \hbox{$\, \buildrel {\scriptstyle <}\over {\scriptstyle\sim}\,$}}
\newcommand{\simgt}{\lower 2pt \hbox{$\, \buildrel {\scriptstyle >}\over {\scriptstyle\sim}\,$}}
\newcommand{\simlt}{\lower 2pt \hbox{$\, \buildrel {\scriptstyle <}\over {\scriptstyle\sim}\,$}}
\newcommand{\msun}{\hbox{${M}_{\odot}$}}

\newcommand{\asca}{{\emph{ASCA}}}
\newcommand{\chandra}{{\emph{Chandra}}}

\newcommand{\xmm}{{\emph{XMM-Newton}}}

\newcommand{\hst}{{\emph{HST}}}

\newcommand{\rosat}{{\emph{ROSAT}}}
\newcommand{\balq}{BAL~quasar}
\newcommand{\balqs}{BAL~quasars}
\newcommand{\pgone}{PG~1115+080}

\newcommand{\aox}{$\alpha_{\rm ox}$}

\newcommand{\apj}{{Astrophys. J.}}

\newcommand{\apjl}{{Astrophys. J. Lett.}}
\newcommand{\aj}{{Astron. J.}}
\newcommand{\mnras}{{Mon. Not. R. Astron. Soc.}}

\newcommand{\pgbal}{PG~2112+059}

\newcommand{\apm}{APM~08279+5255}
\newcommand{\clover}{H~1413+117}

\title{Probing Broad Absorption Line Quasar Outflows: X-ray Insights}

\author{S.C. Gallagher,\address{Center for Space Research,
        Massachusetts Institute of Technology, 77 Massachusetts
        Avenue, Cambridge, MA 02139, USA}
        W.N. Brandt,$^2$ G. Chartas,$^2$ G.P. Garmire,\address{Department of
        Astronomy \& Astrophysics, The Pennsylvania State University,
        525 Davey Laboratory, University Park, PA 16802, USA}
        and R.M. Sambruna\address{Department of Physics and Astronomy
        and School of Computational Sciences, MS 3F3, 4400 University
        Drive, George Mason University, Fairfax, VA 22030}}

\begin{document}

\maketitle
\begin{abstract}
Energetic outflows appear to occur in conjunction with active mass
accretion onto supermassive black holes.  These outflows are most
readily observed in the $\sim10\%$ of quasars with broad absorption
lines, where the observer's line of sight passes through the wind.
Until fairly recently, the paucity of X-ray data from these objects was
notable, but now sensitive hard-band missions such as \chandra\
and \xmm\ are routinely detecting broad absorption line quasars. 
The X-ray regime offers qualitatively new information for
the understanding of these objects, and these new results
must be taken into account in theoretical modeling of quasar winds.
\end{abstract}

\section*{INTRODUCTION}
\label{sec:intro}

Recent studies of the strong correlations between black hole
masses and the properties of their hosts' galactic bulges 
\citep[e.g.,][]{FerMer2000,GebhardtEtal2000} demonstrate clearly the
intimate interrelation between the growth of black holes and
their host galaxies.  This connection implies a physical mechanism for
regulating the coeval development of these structures.
During the bulk of their accretion phases, supermassive black holes
apparently reveal themselves as luminous quasars
\citep[e.g.,][]{YuTre2002}, and thus quasar winds 
are likely to provide a primary source of feedback. 
These winds are directly observed in the population of 
Broad Absorption Line (BAL) quasars, $\sim10\%$ of the quasar
population that exhibit deep, broad absorption lines from high-ionization
UV resonance transitions. Such blueshifted absorption features
are understood to arise along lines of sight which pass through
radiatively driven winds with terminal velocities reaching 0.1--0.3$c$.
These energetic outflows are important components of quasar environments;
mass ejection is apparently fundamentally linked to active mass accretion.
X-rays, generated in the innermost region surrounding accreting black
holes, travel through the nuclear environments to the observer.  X-ray
studies of BAL quasars thus offer a privileged view through the wind.

\begin{figure}[t!]
\centerline{\includegraphics[width=110mm]{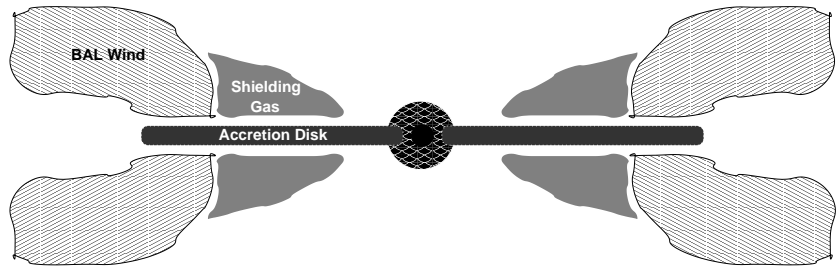}}
\caption{Diagram of a possible geometry for the nucleus of a
  BAL~quasar based on the \citet{MuChGrVo1995} model.  The
patterned gray circle represents the X-ray emitting region surrounding the black hole.
The region of highly ionized plasma marked ``Shielding Gas'' was postulated
 to shield the BAL wind (with a scale of light days) 
from soft X-ray and extreme UV radiation. 
These photons would otherwise strip the
gas in the wind of electrons, thus preventing it from being
radiatively driven by UV line pressure. 
\label{fig:murray}}
\end{figure}

The challenge of accelerating gas to the high velocities observed in
BAL quasars led \citet{MuChGrVo1995} to a picture similar to that presented
in Figure~\ref{fig:murray}.  
In this model, at radii starting at $\sim10^{16}$~cm for a
$10^{8}$~\msun\ black hole,
radiation and gas pressure lift material from the disk photosphere
where it is struck by light from the inner accretion disk and
corona.  The wind, initially
co-rotating with the disk, is accelerated radially by the momentum of
UV line photons; the material launched from the innermost radii
reaches the highest terminal velocities.  In order to launch the wind
at such small distances, a thick layer of ``shielding'' gas, highly ionized plasma
with $N_{\rm H}=10^{22}$--$10^{24}$~\cmsq, was introduced to protect the
wind from becoming completely ionized by soft X-rays.  
In the subsequent hydrodynamical modeling of line-driven quasar winds 
by \citet*{PrStKa2000}, this layer of shielding gas
arose naturally from the simulations.
This scenario produces several predictions for X-ray observations of BAL
quasars:

\begin{itemize}\setlength{\itemsep}{-1mm}
\item{BAL quasars should have intrinsic X-ray
    continuum shapes \citep[with $\Gamma=2.0\pm0.3$; e.g.,][]{GeoEtal2000} and
    spectral energy distributions typical of radio-quiet quasars.}

\item{All BAL quasars should be X-ray weak, i.e., have values of
    \aox\,\simlt$-2.0$,{\footnote{The quantity \aox\ measures the
        relative flux densities at rest-frame 2~keV and 2500~\AA.  For reference,
        a sample of low-redshift radio-quiet quasars without known absorption
        has \aox$=-1.56\pm0.14$ \citep{BrandtEtal2001}; more negative
        values of \aox\ indicate a quasar is X-ray weak relative to its
        UV power.}} 
      as a result of large column densities of
      absorbing gas.  Furthermore, with data of sufficient quality, the spectra will
      show the signatures of absorption by highly ionized gas.}

\item{A correlation between the minimum absorption velocity and \aox\
  is likely to arise due to effects of orientation.  For example, a line of
  sight close to the accretion disk will pass through more X-ray
  absorbing gas, and the minimum observed UV absorption velocity will be close to zero
  as the gas outflow is primarily transverse to the line of sight.}
\end{itemize}

The sample of \balqs\ observed by \chandra\ and
\xmm\ is steadily growing, and the current X-ray data are now of
sufficient quality to investigate
these predictions directly.  In this paper, we summarize the insights offered
from both exploratory and spectroscopic X-ray surveys of \balqs\ to
the understanding of quasar winds. 

\section*{RESULTS FROM EXPLORATORY SURVEYS}
\label{sec:res_exp}

Given the extreme faintness of \balqs\ in the soft X-ray regime
\citep[e.g.,][]{KoTuEs1994,GrMa1996}, obtaining detailed
spectroscopic information requires a significant commitment of
observatory time.  In this situation, only a small fraction of the
population will have data of spectroscopic quality, and there is a
legitimate danger that the conclusions drawn from such limited,
potentially biased samples will not be representative of the
population as a whole.  Pursuing an alternate strategy of exploratory
observations enables much larger numbers of \balqs\ to be observed with
the goal of measuring X-ray flux, hardness ratio, and \aox\ for each object.
The low background and excellent spatial resolution of \chandra\
allow short, exploratory observations of 5--7~ks to reach sensitive
flux limits of $\lesssim10^{-14}$~\flux\ in the 0.5--8.0~keV band 
\citep{GreenEtal2001,GaBrChSa2002}. 

We are in the process of compiling the largest exploratory survey to
date, with a well-defined sample of \balqs\ drawn from the Large
Bright Quasar Survey \citep[LBQS;][]{FoEtal1989,HeFoCh1995}.  
This sample of $z>1.35${\footnote{At $z>1.35$ 
blueshifted \CIV\ absorption, the definitive
\balq\ signature, enters the observed LBQS wavelength range.}} 
\balqs\ is comprised of 22 objects with 
publicly available rest-frame UV
spectroscopy from \citet{WeMoFoHe1991}.   To date, 15 of the 20 objects
observed have been detected, and the remainder have tight upper
limits on both X-ray flux and \aox\ \citep[][in prep.]{GaBrChSa2002}.

\begin{figure}[t!]
  \begin{center}
    \hspace{2mm}
    \begin{minipage}[c]{0.55\linewidth}
      \includegraphics[width=0.72\linewidth]{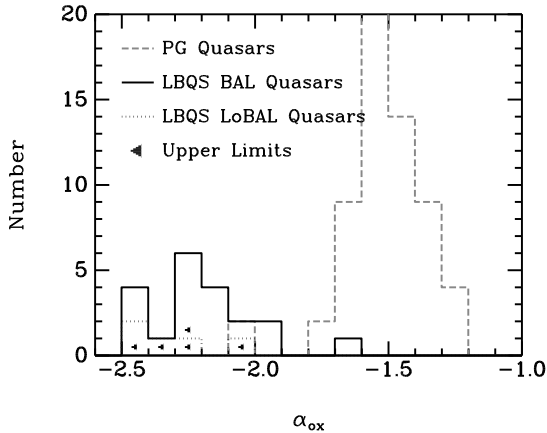}
    \end{minipage}\hfill
    \begin{minipage}[c]{0.40\linewidth}
      \caption{Histogram of the \aox\ values from the LBQS \balq\ exploratory
  survey.  The solid line is the entire sample, while the dotted line
  is the sample of Lo\balqs.  The arrows indicate upper
  limits. For comparison, the dashed gray line represents the
  low-redshift Palomar-Green sample of unabsorbed radio-quiet quasars
  from \citet{BrLaWi2000}.}
    \end{minipage}
  \end{center}
\end{figure}

Though the \balqs\ in our sample are indeed weak in X-rays with a median value of
\aox\,$=-2.2$ (see Figure~2), they are 
generally detectable with the current generation of X-ray observatories.
For those objects that are detected, a measurement of the hardness
ratio{\footnote{The hardness
    ratio is defined as $HR = (h-s)/f$ where $ h=2$--8~keV
counts,  $s=0.5$--2.0~keV counts, and $f=0.5$--8.0~keV counts.}}
 indicates what fraction of the full-band (0.5--8.0~keV) counts are coming
from hard band (2--8~keV) X-rays.  Absorbed quasar spectra will have
larger values of the hardness ratio as
the cross-section for photoelectric absorption in the X-ray band 
decreases rapidly with increasing energy.
The trend of increasing hardness ratio with decreasing \aox\
illustrated in Figure~3 is therefore
consistent with the understanding that X-ray faintness arises from
intrinsic absorption.   While this result is not new, it supports
the idea that \aox\ can be used in a rough sense to indicate the extent of
absorption.  Following this premise, \aox\ might be expected to
correlate with UV absorption properties such as \CIV\
absorption-line equivalent width, as demonstrated 
by \citet{BrLaWi2000} for their sample of Palomar-Green quasars. 
To complement \citet{BrLaWi2000} we are probing the extreme end of
this parameter space with the \balqs.
In this regime, there are no apparent correlations between \aox\ and
several absorption-line properties for the \balqs\ , most
significantly the minimum velocity of \CIV\ absorption shown in Figure~3.  

Furthermore, the large column densities of absorbing
gas implied by values of \aox\,$\lesssim-2.0$ coupled with the blue
UV continua of the majority of our sample imply that the
X-ray absorbing gas contains little dust, at least as we know it.
This is consistent with the X-ray absorber being located within the
dust sublimation radius. 
Alternatively, the bulk of the X-ray absorption may lie interior to
the UV emission region, though this implies even smaller radii for
the absorption.

We also confirm the result of \citet{GreenEtal2001} that the
\balqs\ with broad \MgII\ absorption, the Lo\balqs,  are notably X-ray weaker than
those with only high-ionization BALs.  In fact, only one of the
Lo\balqs\ in our sample was actually detected (see
Figure~2).  Given that Lo\balqs\ generally have reddened
optical and UV continua, it is unclear whether the extreme
X-ray absorption that they suffer is occuring on the small scales
inferred for the high-ionization \balqs.  However, X-ray variability 
of the Lo\balq\ Mrk~231 suggests that a substantial X-ray absorber 
is located within $\sim10^{15}$~cm of the X-ray continuum source \citep{GaEtal2002b}.

\begin{figure}[t!]
\hspace{8mm}
\includegraphics[width=63mm]{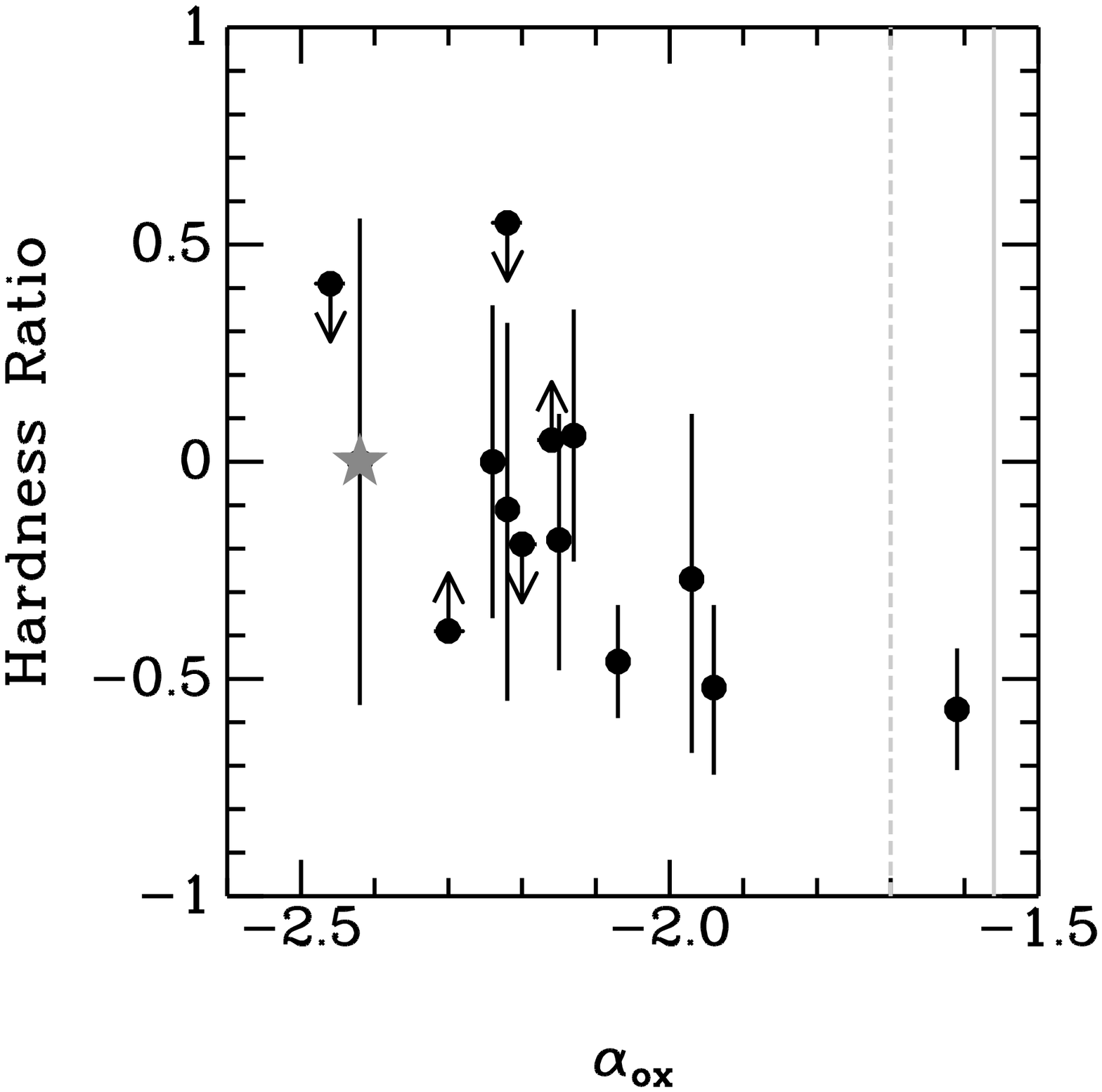}
\hspace{10mm}
\includegraphics[width=63mm]{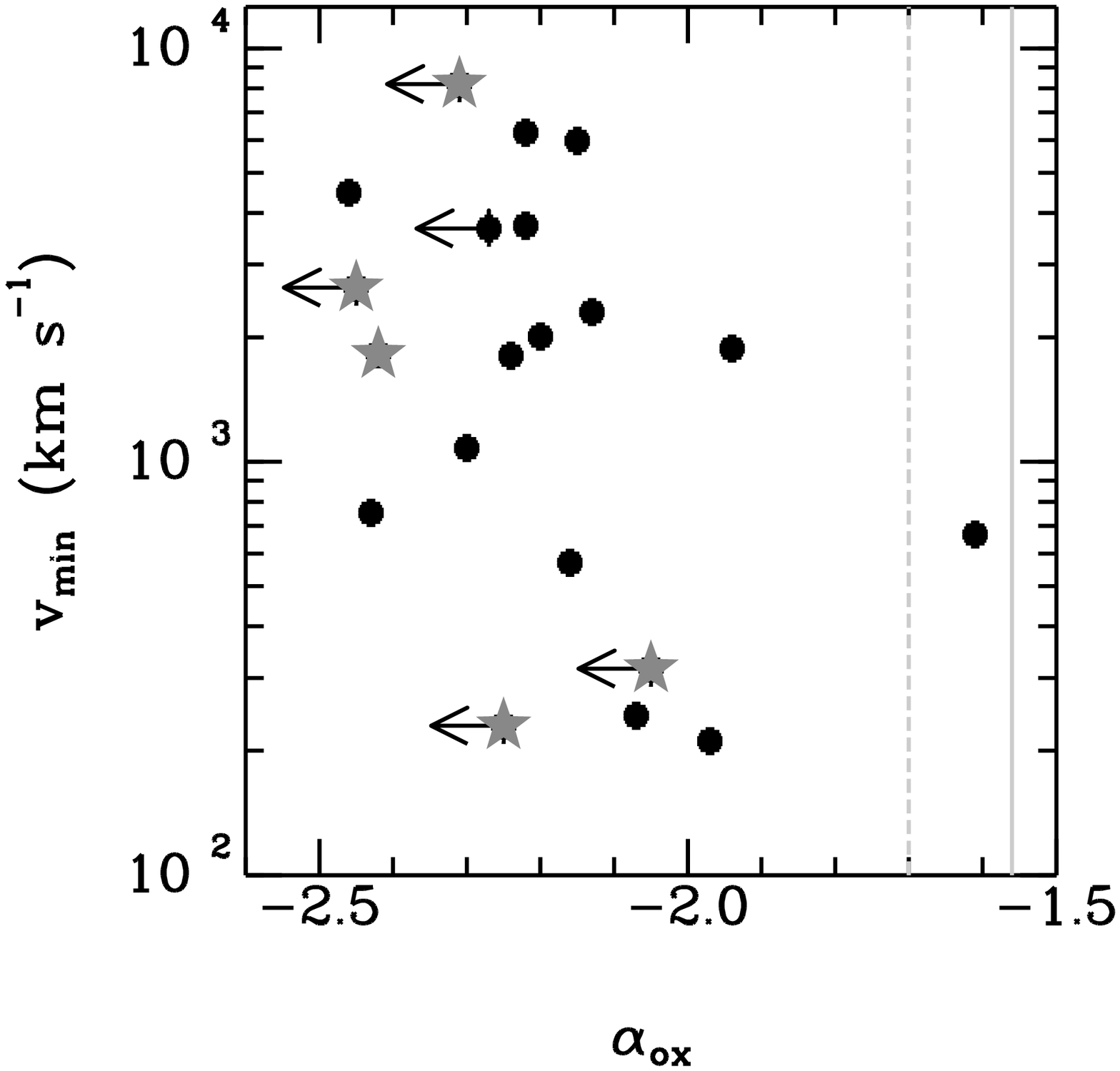}
\vspace*{-12mm}
\caption {\textbf{Left panel:} Hardness ratio versus \aox.
  The trend of increasing hardness ratio with decreasing \aox\
  is consistent with absorption as the primary cause of X-ray weakness
  in \balqs.  \textbf{Right panel:}   
  Minimum velocity of \CIV\  absorption versus
  \aox.  Contrary to expectations from the shielding gas models
  \citep[$\S$5,][]{MuChGrVo1995}, there
  is no apparent correlation between these parameters. 
\label{fig:aox2}}
\end{figure}

\section*{RESULTS FROM X-RAY SPECTROSCOPY}

Though exploratory surveys are essential for determining the general
X-ray properties of the population as a whole, spectroscopic
investigations are also required to obtain more detailed information
on the X-ray continuum shape and absorption properties.
\citet{GaBrChGa2002} gathered together 8 spectroscopic-quality
observations to investigate general trends from these data.  In
general, they found that the underlying X-ray continua are consistent
with normal radio-quiet quasars, with significant, complex absorption.
Possible sources of the observed complexity (illustrated in Figure~4) 
include partial covering of the X-ray continuum, ionized gas, and
velocity structure similar to that seen in the UV regime.  
Furthermore, normalizing the power-law continua above 5~keV to correct
for absorption resulted in typical radio-quiet quasar values of \aox.
All of these results are consistent with the \citet{MuChGrVo1995} picture.

\begin{figure}
  \begin{center}
    \begin{minipage}[c]{0.50\linewidth}
      \includegraphics[width=0.63\linewidth,angle=-90]{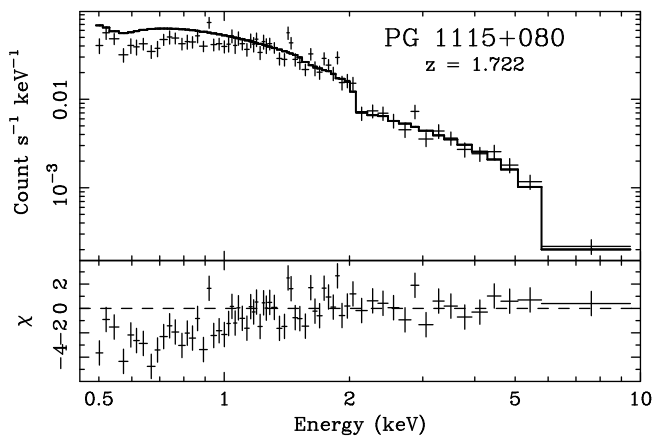}
    \end{minipage}\hfill
    \begin{minipage}[c]{0.40\linewidth}
      \caption{Observed-frame ACIS-S3 spectrum of PG~1115+080.  This
  spectrum has been fit with a power-law model above rest-frame 5~keV
  which has then been extrapolated back to lower energies.   The
  significant negative residuals below $\sim1.2$~keV indicate strong
  absorption, and the structure rules out simple absorption by neutral gas. 
  While the nature of this complex absorber is
  not well-constrained, it is consistent with a high ionization parameter.}
    \end{minipage}
  \end{center}
\end{figure}

However, some unexpected results have also arisen.  Among the most
interesting is the observation of two absorption lines in the spectrum of
\apm, apparently from highly ionized Fe 
outflowing with speeds of $\sim0.2$ and $\sim0.4c$ \citep{ChBrGaGa2002}. 
In radiatively driven flows, 
such high terminal velocities occur in gas launched at much
smaller radii than the UV BAL wind, which has
terminal velocities of $\sim12,000$~\kms.  In addition, 
when compared with an \xmm\ observation of this object taken 
$\sim12$~rest-frame days later \citep{HaScKo2002}, there is a clear
indication of variability in the absorber properties. 
\apm\ is not unique; similar absorption lines have also been seen in
\pgone\ (Chartas et al., in prep.).
These results are challenging for models in which the X-ray 
and UV absorption arise in the same medium and are
predicted to have the same velocity structure \citep[e.g.,][]{Elvis2000}.
Significant iron emission has also been observed in some
objects including \clover\ \citep{OshimaEtal2001,GaBrChGa2002}.

In another unanticipated result, the formerly X-ray brightest
\balq\ \pgbal\ \citep{GaEtal2001a} exhibited significant variability
between the \asca\ observation of
Oct 1999 and the \chandra\ observation of Sep 2002.  In the most
recent \chandra\ data, the power-law continuum normalization dropped
by a factor of a few while the measured column density of the absorber
increased significantly. Notably, concurrent \hst\ STIS data
of the \CIV\ absorption line showed no large changes from earlier
UV observations.  For \pgbal\ at least, evidence for distinct
X-ray and UV absorbers is mounting.

\section*{IMPLICATIONS FOR THEORETICAL MODELS OF QUASAR OUTFLOWS}

In general, X-ray studies are providing support for the family of radiatively
driven wind models.  \balqs\ are generally
X-ray weak with values of \aox\ indicating a significant reduction in
observed X-ray flux relative to the UV continuum. 
When sufficient data are available, the intrinsic X-ray power, corrected
for absorption, is typical of  normal radio-quiet
quasars.  The complexity in the X-ray absorbers is consistent with highly
ionized gas as predicted for the shielding gas;
however the ionization state cannot be constrained given the current
data quality.

Though the wind models have had some success, 
the lack of any apparent correlation between the UV
absorption-line and X-ray properties is contrary to the model predictions.
While certainly X-ray and UV absorption are found together in
the same objects, the connection between the two is not straightforward.  
This is perhaps not unexpected given the complexity of the UV
BALs and potential quasar-to-quasar variations in wind properties.
The assertion of \citet{Elvis2000} that the
UV and X-ray absorption arise from the same medium is not
easily reconciled with the lack of connection between X-ray
and UV variability seen in \pgbal\ and the mismatch in X-ray and UV absorption
properties seen in \apm.  In addition, the large-velocity outflows
seen in \apm, which 
imply a much smaller launching radius that for the BAL gas, do
not fit within the \citet{Elvis2000} picture with a narrow range of
launching radii for the outflow.  

One of the problems for theoretical wind models with small launching
radii is the lack of variability in velocity structure in the UV absorption
lines on observed-frame timescales of years \citep[e.g.,][]{deK1997}.
However, the most comprehensive absorption-line variability investigation to
date, the 3~yr study of \citet{Barlow1993}, was not long enough for a
conclusive test; the hydrodynamic wind models of \citet{PrStKa2000} generated
instabilities on rest-frame timescales of $\sim3$~yr.
With longer time baselines, velocity increases have been seen for at least
one object \citep{VilIrw2001}, and an extended study is certainly
warranted to investigate this issue further.  
If the bulk of the X-ray absorbing gas is on smaller scales than the
BAL gas, it is also possible that X-ray absorption variability on
shorter timescales may regularly occur, as seen for \pgbal\ and \apm.  
To make accurate predictions for such events, the
hydrodynamic models may need to be extended to smaller radii
($<10^{15}$~cm) with higher resolution, 
and also more explicitly include the effects of
Compton pressure. For highly ionized gas, this will dominate the
radiation force (D. Chelouche, priv. comm.).
Additional items for an X-ray observer's wish list from theoretical
modeling are specific predictions for the strength and profiles of Fe emission and absorption lines.

An outstanding issue for future investigations concerns the velocity of the X-ray 
absorbing gas in most \balqs; is it stalled, outflowing, or inflowing?  
\citet{PrStKa2000} claimed that highly ionized gas should be inflowing or stalled, 
but the evidence from \apm\ and \pgone\ suggests otherwise.  
An X-ray gratings observation of a BAL quasar offers the potential for
a qualitative advance in this field.  The improved resolution of
gratings data, from $\Delta v\sim20,000$ \kms\ for CCD data to
$<1300$~\kms, would 
enable a more penetrating and subtle investigation into the connection
between UV and X-ray absorption. 
In particular, significant constraints could be placed
on the velocity, column density, and ionization state of the X-ray
absorbing gas.  These values can then provide information on the
location of the X-ray absorber and the mass-outflow rate,
parameters of fundamental physical importance.

\section*{ACKNOWLEDGEMENTS}
We thank Sera Markov, Bev Wills, and Doron Chelouche for helpful discussions.
This work was made possible by \chandra\ X-ray Center grants GO1-2105X and
GO2-3129A as well as NASA grant NAS 8-38252 for the ACIS instrument team.


\medskip
\par\noindent
Email address of S.C. Gallagher: {\em scg@space.mit.edu}.

\end{document}